\def\year{2022}\relax
\documentclass[letterpaper]{article} 
\usepackage{aaai22}  
\usepackage{times}  
\usepackage{helvet}  
\usepackage{courier}  
\usepackage{multirow}
\usepackage[hyphens]{url}  
\usepackage{graphicx} 
\usepackage{natbib}  
\usepackage{caption} 
\DeclareCaptionStyle{ruled}{labelfont=normalfont,labelsep=colon,strut=off} 
\usepackage{booktabs}
\usepackage{color}

\title{Characterizing Information Shared by Participants to Coding Challenges: The Case of Advent of Code}
\author{
    Francesco Cauteruccio\textsuperscript{\rm 1}\equalcontrib,
    Enrico Corradini\textsuperscript{\rm 2}\equalcontrib,
    Luca Virgili\textsuperscript{\rm 2}\equalcontrib
}
\affiliations{
    \textsuperscript{\rm 1}DIEM, University of Salerno\\
    Fisciano, I84084, Italy\\
    fcauteruccio@unisa.it,\\
    \textsuperscript{\rm 2}DII, Polytechnic University of Marche\\
    Ancona, I60131, Italy\\
    e.corradini@univpm.it, luca.virgili@univpm.it
}

\begin{document}

\maketitle

\begin{abstract}

Advent of Code (AoC from now on) is a popular coding challenge requiring to solve programming puzzles for a variety of skill sets and levels. AoC follows the advent calendar, therefore it is an annual challenge that lasts for 25 days. AoC participants usually post their solutions on social networks and discuss them online. These challenges are interesting to study since they could highlight the adoption of new tools, the evolution of the developer community, or the technological requirements of well-known companies. For these reasons, we first create a dataset of the 2019-2021 AoC editions containing the discussion threads made on the subreddit {\tt /r/adventofcode}. Then, we propose a model based on stream graphs to best study this context, where we represent its most important actors through time: participants, comments, and programming languages. Thanks to our model, we investigate user participation, adoption of new programming languages during a challenge and between two of them, and resiliency of programming languages based on a Stack Overflow survey. We find that the top-used programming languages are almost the same in the three years, pointing out their importance. Moreover, participants tend to keep the same programming language for the whole challenge, while the ones attending two AoCs usually change it in the next one. Finally, we observe interesting results about the programming languages that are ``Popular'' or ``Loved'' according to the Stack Overflow survey. Firstly, these are the ones adopted for the longest time in an AoC edition, thanks to which users have a high chance of reaching the end of the challenge. Secondly, they are the most chosen when a participant decides to change programming language during the same challenge.

\end{abstract}

\section{Introduction}


In the last years, coding challenges took on a fundamental role in teaching and learning computer science. Although coding challenges have been around for a significant amount of time, thanks to forums, messaging systems, and popular social media, it is now extremely easy to take part in one of them. This kind of challenges can have different purposes, ranging from teaching a particular subject via gamification approaches to even use them to assess the skills of hypothetical candidates for a job position~\cite{peterson2020coding,wyrich2019theory}. One of the most interesting coding challenges in the computer science world is Advent of Code (AoC). This challenge mimics an advent calendar: it starts on December, 1st and ends on December, 25th. Each day, a disguised coding puzzle is given to participants, whose aim is to code a solution in any language of preference, and submit the output they obtain by applying this solution to the input they receive from the puzzle. In this context, an interesting aspect to study is the information shared by the participants, as well as the interactions between them. These aspects are of uttermost importance from different points of view. As an example, interactions can provide insights into the problem-solving strategies and techniques used by coders, as well as how they collaborate and share information between them. Also, participants providing their solutions can fuel discussions on how they designed them and whether a specific background is needed in order to do that. Coding challenges often provide a way to connect participants, such as a discussion place or a forum, where they can exchange their thoughts. This is not the case for AoC. Nevertheless, they gather up on Reddit, and in particular on the {\tt /r/adventofcode} subreddit, to discuss the challenge and show their solutions. Therefore, we target this subreddit as our data source for analyzing the interactions between the participants.

To characterize and investigate the information shared by the participants and their interactions, in this paper, we focus on extracting and analyzing the discussions produced by AoC participants on Reddit. In particular, the contribution of our work is two-fold. First off, we start by gathering and aggregating information posted by participants of the AoC coding challenge on Reddit. This step enables to study and analyze information posted by participants, which could not be possible otherwise. We focus on the AoC 2019, 2020, and 2021 editions, by analyzing over 23K comments, including more than 5K unique participants. We also represent these data via the stream graph model. Thanks to its flexibility, this model allows for a series of analyses which we encompass in the following research questions:

\begin{itemize}
    \item \textbf{RQ1: Analysis of user participation in AoC} – How can participation to the challenge be characterized? We conduct several analyses to investigate users’ participation in the challenge. In particular, we focus on the characterization of their temporal distribution within the challenges, and their programming languages, by quantifying them and discussing their choices.
    \item \textbf{RQ2: Adoption of new programming languages} – Does this challenge motivate users to employ new programming languages to solve puzzles? We answer this question by characterizing participants who change languages within the same challenge and between two of them. In particular, we analyze intra-challenge changes, i.e., participants switch their language of choice during the challenge, and inter-challenge changes, i.e., they swap languages between years.
    \item \textbf{RQ3: Programming language resiliency} – What is the time span of a programming language in this kind of challenge? How much a programming language is attractive? We investigate the resiliency of programming languages by studying their adoption in the challenge. We also study their resilience w.r.t. to a classification of programming languages based on different features.
\end{itemize}

We find several insights through these research questions, ranging from the ones with expected answers to the ones which are useful to shed light on interesting aspects of coding challenges. For instance, we find that the mean number of participants decreases as the difficulty of the challenge increases, which is somewhat of an expected result. Also, we find that participants that attend two AoC editions tend to use different programming languages each time, suggesting that this coding challenge can motivate participants to try out different programming languages.

The remainder of this paper is as follows. We first provide a background on the AoC coding challenge and its importance. Then, we introduce the dataset, describe the process to identify programming languages, and present the context modeling. We continue by presenting our research questions along with the obtained results and a discussion for each of them. Afterward, we illustrate the related literature. Finally, we draw our conclusion and look at possible future developments.


\section{Advent of Code}
\label{sec:AoC}

Advent of Code is a popular coding challenge in the programming and coding world. By quoting from its website\footnote{\url{https://adventofcode.com}}, ``AoC is an Advent calendar of small programming puzzles for a variety of skill sets and skill levels that can be solved in any programming language of preference''. The challenge can be explained as follows. Since it mimics an Advent calendar, the challenge takes place each year in December starting on the 1st and ending on the 25th. On a daily schedule, a puzzle is proposed to the participant. This is always a coding problem in disguise. It consists of a fictional backstory that is the same for all participants but each participant receives a different piece of input data. The participant should solve the problem by coding a solution in any programming language of preference, and then generate a result according to the received input data. Eventually, the participant submits the result to the AoC website, which returns whether the result is correct for that day. Each puzzle consists of two challenges, which are often correlated. The participant must solve the first challenge in order to unlock the second one. There is a global leaderboard, in which participants are ranked based on the time of completion of the daily riddle. Each participant can decide whether to compete on the global leaderboard or not.
Advent of Code is a popular coding challenge in the programming and coding world. By quoting from its website\footnote{\url{https://adventofcode.com}}, ``AoC is an Advent calendar of small programming puzzles for a variety of skill sets and skill levels that can be solved in any programming language of preference''. The challenge can be explained as follows. Since it mimics an Advent calendar, the challenge takes place each year in December starting on the 1st and ending on the 25th. On a daily schedule, a puzzle is proposed to the participant. This is always a coding problem in disguise. It consists of a fictional backstory that is the same for all participants but each participant receives a different piece of input data. The participant should solve the problem by coding a solution in any programming language of preference, and then generate a result according to the received input data. Eventually, the participant submits the result to the AoC website, which returns whether the result is correct for that day. Each puzzle consists of two challenges, which are often correlated. The participant must solve the first challenge in order to unlock the second one. There is a global leaderboard, in which participants are ranked based on the time of completion of the daily puzzle. Each participant can decide whether to compete on the global leaderboard or not.

AoC was launched on December 1, 2015, and by the end of the event, a grand total of 52,000 participants was collected. In the following years, AoC started reaching higher ceilings in the number of participants. As an example, during the COVID-19 pandemic, the event saw a 50\% spike in traffic, with more than 180,000 participants worldwide \cite{aoc-numbers}. Several aspects could have fueled the success of AoC. For instance, the challenges do not impose a time limit. In this sense, participants can take part in it at their own pace without the urge to complete a task. It also means that participants are more likely to compete in AoC using a programming language they are not proficient with. It is known that previous knowledge of a language could potentially interfere with learning a new one~\cite{shrestha2020here}, therefore approaching AoC with the mindset of learning a new programming language seems a suitable scenario. Moreover, there are no particular constraints on the required background; puzzles do not always require a profound knowledge of Computer Science topics, and most of them only require problem-solving skills. This aspect reinforces the observation that AoC could be used as a ``sandbox'' for learning new programming languages, which in turn becomes an interesting aspect to study.

\section{Materials and Methodology}
\label{sec:Materials}

In this section, we introduce the dataset that we used to carry out our analyses and the model to deal with our scenario. Specifically, we start by explaining the details about the creation of our dataset along with some descriptive statistics. Then, we show the process to extract the used programming language from a comment reporting the solution proposed by a user of a puzzle. Finally, we describe the stream graph model that we used to represent the AoC scenario.

\subsection{Dataset Description}
\label{sub:Dataset-Description}

In order to study Advent of Code challenges, we need a dataset containing the solutions proposed by the participants for each day's puzzle. To this end, we monitored the {\tt /r/adventofcode} community on Reddit \cite{cauteruccio2022extraction}. Specifically, a moderator of the community makes a ``megathread'' post for each day of an AoC edition in order to give the possibility to the other participants to post the solution for the corresponding puzzle. Megathreads are discussions referring to a particular context; these provide a unique opportunity to perform topic-specific analyses~\cite{basile2021dramatic}. In each megathread post, we will find many comments containing a code implementing the solution for the daily puzzle. Then, other users can further comment on that solution giving tips or asking for explanations. In this way, users give rise to a series of comments, where the first one concerns the solution proposed by a participant, and the next ones discuss it. A participant can propose more than one solution. 

\begin{figure}[!ht]
\centering
\includegraphics[width=0.4\textwidth]{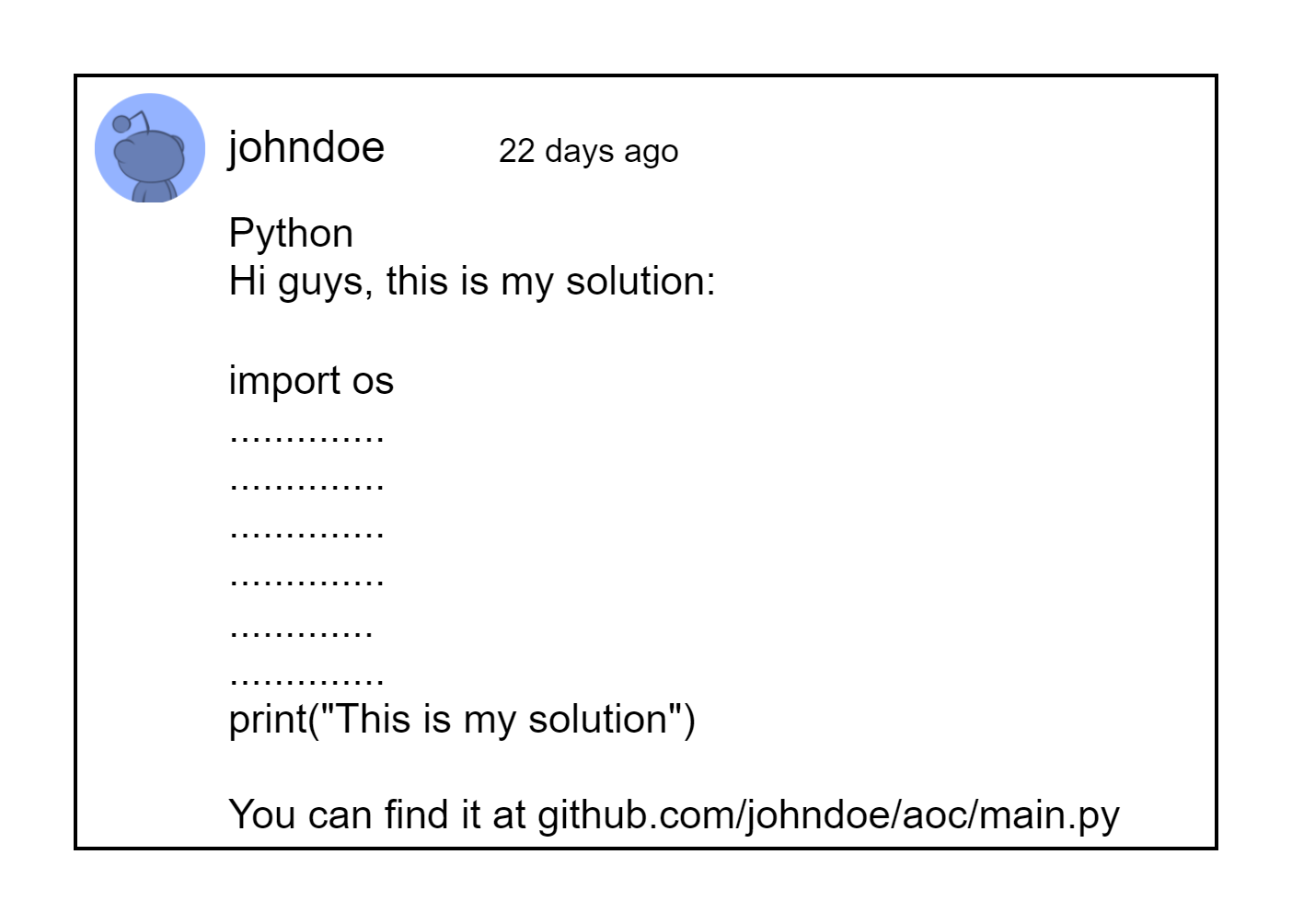}
\caption{Schematic representation of the common structure of a comment in the AoC ``megathread''}
\label{fig:Comment-Example}
\end{figure}

Thanks to Reddit API, we downloaded all the megathread posts of the 2019-2021 AoC editions along with the corresponding comments. We kept only the first comment of the corresponding series of comments which contained the solution of the participant. We report a schematic representation of the common structure of a comment in Figure \ref{fig:Comment-Example}.

From Figure \ref{fig:Comment-Example}, we can see two important elements useful to identify the programming language in a comment. The participant usually writes the programming language as the first line in the comment, and sometimes he/she adds the corresponding script in an online repository, e.g., GitHub. As will be clear in the following section, these two elements will be used to extract the programming language of participant comments.

Given our dataset of comments presented above, along with the extracted programming languages, we obtained everything needed for our analyses. Some dataset statistics are reported in Table \ref{tab:Dataset-Statistics}.

\begin{table}[!ht]
\centering
\begin{tabular}{l|l}
\toprule
No. of AoC editions & 3 \\ 
No. of megathread & 75 \\
No. of comments & 24,998 \\
No. of participants & 5,398 \\
No. of programming languages & 118 
\\ \bottomrule
\end{tabular}
\caption{Dataset descriptive statistics}
\label{tab:Dataset-Statistics}
\end{table}

From the analysis of Table \ref{tab:Dataset-Statistics}, we observe that our dataset contains the AoC editions of the 2019-2021 years. Therefore, we have 75 megathread posts (25 for each AoC edition), and 24,998 comments for all AoC editions. The unique number of participants is 5,398, and some of them have attended more than one AoC edition. Finally, we extracted 118 different programming languages from all the comments, which points out a very high diversity of tools used in these challenges.

\subsection{Programming Languages Identification}
\label{sub:Programming-Identification}

As previously stated, our dataset is composed of comments of users participating in the AoC challenge. Starting from comment raw texts, we must define a way to extract the programming language used for the solution of a puzzle. Given the homogeneous structure of comments reported in Figure \ref{fig:Comment-Example}, we designed a procedure to get the corresponding programming language. We implemented it as a Python script. Starting from a comment, our script performs four sequential steps that are designed to look at different parts of the text in order to extract the programming language. If it is found in one of those steps, the others are not executed.

The first step of our script checks if the comment is marked with the name of the programming language it refers to, which could be something like ``Python'' or ``[Python]'', such as in Figure \ref{fig:Comment-Example}. The second one searches for the name of the programming language in the comment text. The third step verifies if the user reported a link to a file with an extension related to a programming language (such as ``.py'' for Python, ``.kt'' for Kotlin, etc.). If all of them fail, the fourth step uses Part-of-Speech tagging to identify the proper names in the comment and verifies if one of them is a programming language. Keeping in mind that text mining is not the scope of this work, these four steps cover all the possible ways a programming language can be written in a comment.


Thanks to this procedure, we were able to identify the programming languages used in the participants' comments and we can now model the AoC scenario in order to perform our analyses.

\subsection{Context Modeling}
\label{sub:Model}

We used stream graphs to model our scenario \cite{latapy2018stream}. We recall that a stream graph is defined as $S=(T,V,W,E)$. Here, $V$ is a finite set of nodes, $T$ is a measurable set of time instants, $W \subseteq T \times V$ is a set of temporal nodes, and $E \subseteq T \times V \otimes V$, such that $(t,uv) \in E$ implies $(t,u) \in W$ and $(t,v) \in W$.

In order to model our context of analysis, we are going to add another set to the previous definition. We extend it as $S = (T,V,W,E,L)$, where $L$ is the set of edge labels. While $T$, $V$, and $W$ remain unchanged from the previous definition, the set of edges changes because an edge have also a label in our model. So, we have $E \subseteq T \times 2^V \times L$. It is a subset of all the possible couples of nodes for each temporal instant, with labels. An edge is defined as $(t,uv,l)$, where $t \in T$ is the time instant in which the edge exists, $u,v \in V$ are the nodes linked together and $l \in L$ is the label. In our context, a node is a user and a label is a programming language. Therefore, an edge indicates a solution using a certain programming language posted by a user in a discussion. A time instant $t$ can be a day of the challenge or the year of an AoC edition. In the following, we will use $d$ for days and $y$ for years, to distinguish them. In addition, we will use $D$ for the set of days and $Y$ for the set of years, where $T = D \cup Y$.

We also introduce a function $\gamma: V \to C$, which gives the set $C$ of comments written by a set of users.

We will adopt the same notation introduced in \cite{latapy2018stream}. Some examples used in this work are the following:
\begin{itemize}
    \item $V_t$, the set of nodes we find at the instant $t$;
    \item $\gamma(V')$, the function that gives the set of comments $C_{V'}$ written by a set of users $V' \subseteq V$; $\gamma_t(V')$ gives the set of comments at time $t$;
    \item $\delta(V')$, the function that gives the set of programming languages $L_{V'}$ used by the set of users $V' \subseteq V$; $\delta_t(V')$ gives the set of programming languages at time $t$;
    \item $E_{l}$, the set of edges with label $l$; $E_{lt}$ at time $t$;
    \item $L_t$, the set of programming languages we find at time instant $t$;
    \item $\bar{v}_d = \sum_{\forall y \in Y}V_{y_{d}}/|Y|$, the mean number of users between the three editions considered that we have for a day $d$ of a challenge;
    \item $A_{vl} = \{t_1..t_n : \forall t \in T, \exists \ t_1..t_n \in T, l \in \delta_t(v), t_1 \leq t \leq t_n\}$, the set of sequences of consecutive time instants $t_1..t_n$ in which the participant $v$ uses the same programming language $l$; $max(A_{vl})$ is the longest sequence in the set.

\end{itemize}
We reported the main notations used in the following. We decided to model our context via a stream graph thanks to its flexibility: indeed, the model is adaptable to any analysis needed and enables several views of the data.

\section{Results and Discussion}
\label{sec:Results}

In this section, we seek to answer the three research questions (RQ) presented in the Introduction. Regarding the first one, we analyze user participation. For the second one, we study the trend of adoption of new programming languages, and finally, for the third one, we characterize the programming languages throughout the AoC editions.

\subsection{RQ1: Analysis of user participation in AoC}
\label{sub:RQ1}

The first RQ we want to address regards the analysis of user participation in AoC during the considered editions. As a first step, in Table~\ref{tab:Overview-Challenges}, we report the number of participants $|V_y|$, the number of comments $|C_y| = |\gamma_y(V)|$, and the number of the different programming languages $|L_y|$, for each year $y \in Y$.

\begin{table}[!ht]
\centering
\begin{tabular}{c|c|c|c}
\toprule
$y$ & $|V_y|$ & $|C_y|$ & $|L_y|$ \\ \midrule
2019 & 1,096 & 3,374 & 77 \\ 
2020 & 2,485 & 9,979 & 91 \\ 
2021 & 2,876 & 11,645 & 92 
\\ \bottomrule
\end{tabular}
\caption{Challenge participants, comments, and unique programming languages throughout the AoC editions}
\label{tab:Overview-Challenges}
\end{table}

From Table \ref{tab:Overview-Challenges}, we observe that the number of participants in 2021 is almost three times the one in 2019, while the number of comments has grown more than three times in the same period. This result highlights the constant growth of interest of the online community in the AoC challenges. Moreover, the number of unique programming languages went from 77 in 2019 to 92 in 2021. Along with a higher number of participants, users are testing many different languages. As we noted above, participants could approach AoC in order to improve with or learn a new programming language. We study this aspect in the next section.

\begin{figure}[!ht]
\centering
\includegraphics[width=0.45\textwidth]{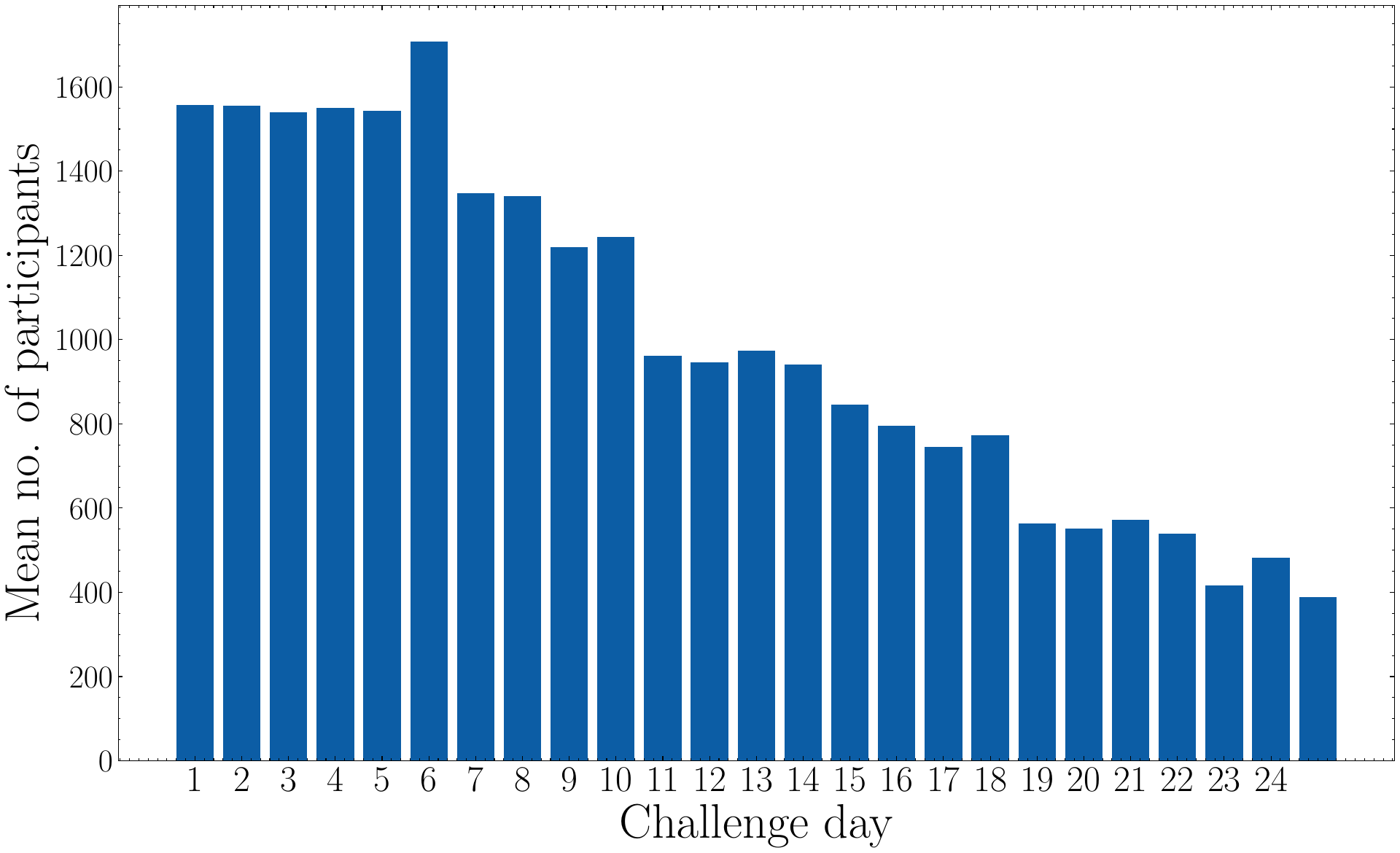}
\caption{Mean number of participants $\bar{v}_d$ during the 2019-2021 AoC editions for each day}
\label{fig:Challenge-Participants}
\end{figure}

Since puzzles during an AoC edition become more complicated as the challenge advances, in Figure \ref{fig:Challenge-Participants}, we report the mean number of participants $\bar{v}_d$ during the 2019-2021 AoC editions for each challenge day. 

From Figure~\ref{fig:Challenge-Participants}, we observe an expected behavior. Indeed, we have many participants in the first 6 days, and then we see a downward trend. On the last days, we have less than one-third of the participants than the first day. However, this is expected since the puzzles become more complex, and participants may find difficulties in solving them.

\begin{figure}[!ht]
\centering
\includegraphics[width=0.45\textwidth]{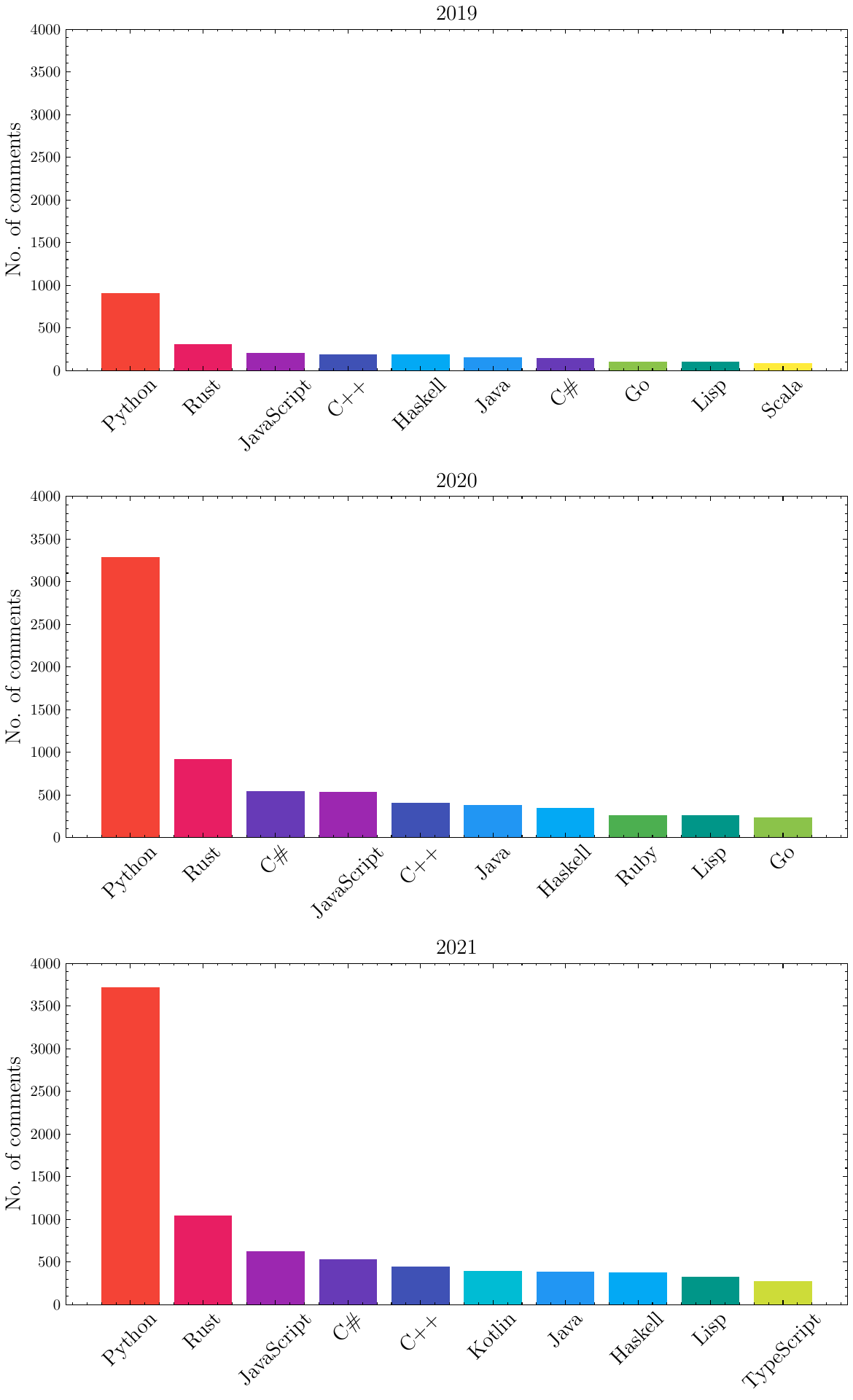}
\caption{Most used programming languages throughout the 2019-2021 AoC editions}
\label{fig:Most-Used-Programming}
\end{figure}

One aspect that is highly correlated with user participation in an AoC edition is the used programming languages. In Figure \ref{fig:Most-Used-Programming}, we report the top 10 most used programming languages appearing in $|L_y|$ for each year $y$.

From Figure \ref{fig:Most-Used-Programming}, we can extract some insights. First of all, we observe that 8 out of 10 of the top programming languages are the same throughout the 2019-2021 AoC editions. Specifically, the top 2 languages are always Python and Rust. We also note that the number of comments increased a lot from 2019 to 2021 for each programming language, which is a trend we highlighted previously. Finally, we see that Python is the most used programming language. It is probably an expected result since Python is very versatile and easy to use, and nowadays it is employed in many different tasks ranging such as web development, blockchain, data analysis, and artificial intelligence. Since its characteristics and its heavy usage in industry, it is reasonable that Python is one of the most used languages in an AoC edition.

\begin{figure}[!ht]
\centering
\includegraphics[width=0.45\textwidth]{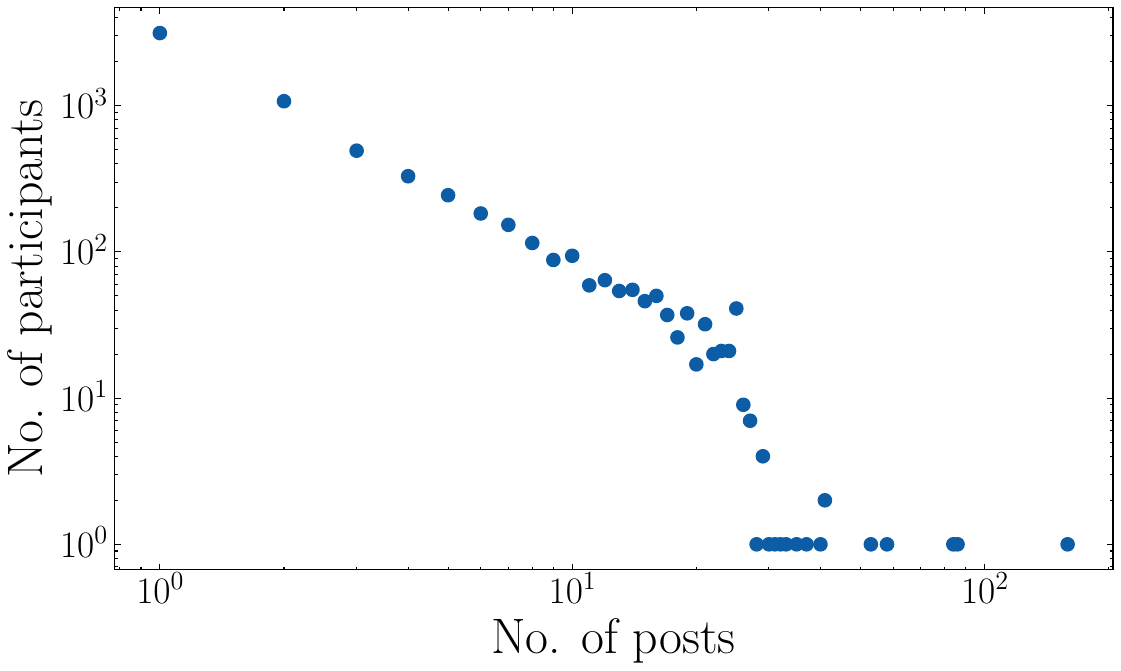}
\caption{Distribution of the number of participants $|V|$ against the number of comments they made $|\gamma(V)|$}
\label{fig:Participants-Comments}
\end{figure}

Finally, in order to analyze another aspect of user participation, we compute the distribution of the number of participants $|V|$ against the number of comments they made $|\gamma(V)|$ throughout the 2019-2021 AoC editions in Figure \ref{fig:Participants-Comments}.

From the analysis of Figure \ref{fig:Participants-Comments}, we observe that the distribution follows a power law, which is in line with other studies in this context \cite{meyerovich2013empirical}, and that it is comparable with Figure \ref{fig:Challenge-Participants}. Indeed, there are lots of participants that write very few comments, while there are very few participants that write lots of comments. One possible reason behind this phenomenon lies in the programming language choice and the increasing complexity of puzzles during the course of an AoC edition. Indeed, as we note in Figure \ref{fig:Challenge-Participants}, participants tend to not participate as the challenge advances, which means that fewer users provide solutions during the last days. Similarly, some users test new programming languages at the beginning of an edition, which is itself a challenge. Many of them cannot keep up with the increasingly complex puzzles and therefore do not contribute anymore.

\subsection{RQ2: Adoption of new programming languages}
\label{sub:RQ2}

\begin{figure}[!ht]
\centering
\includegraphics[width=0.45\textwidth]{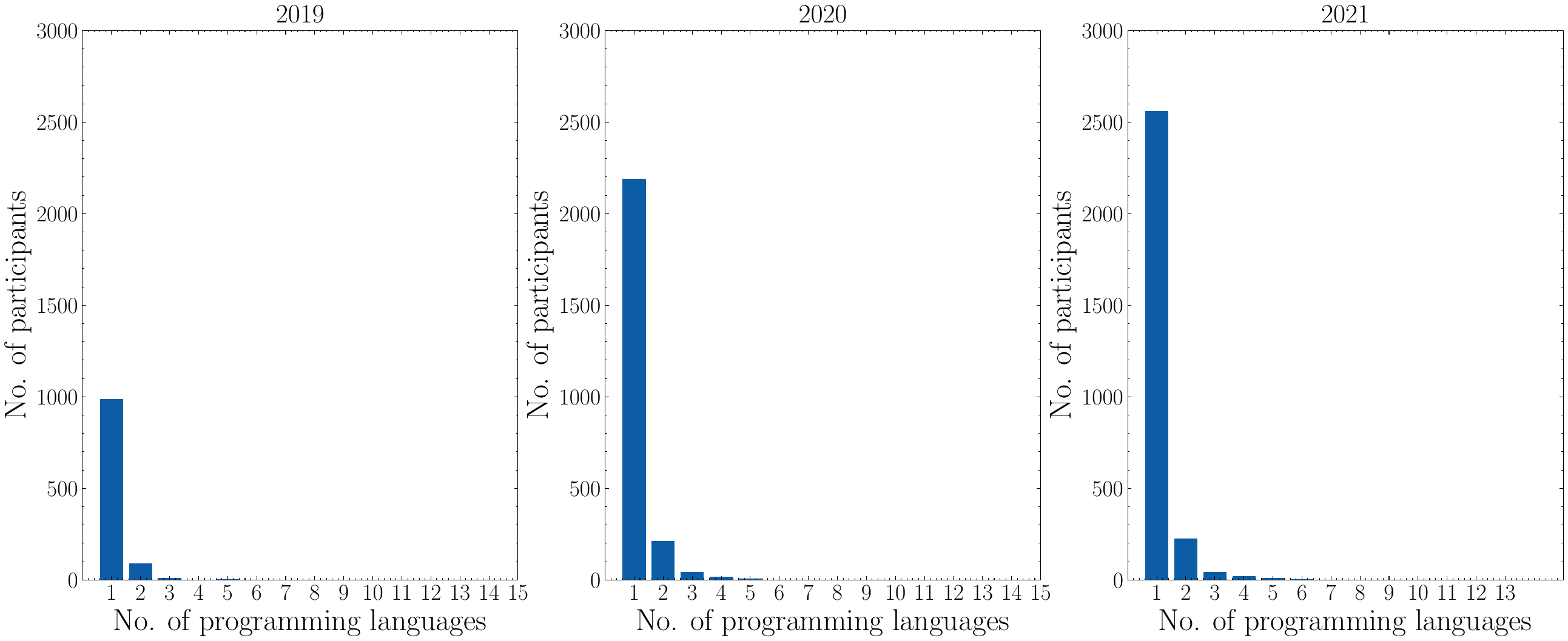}
\caption{Number of programming languages used by the participants throughout the 2019-2021 AoC editions}
\label{fig:Language-Participants}
\end{figure}

In this section, we investigate if users approach AoC with the mindset of learning a new programming language or using different ones. To this end, we start by computing the distributions of the number of participants $|V_y|$ against the number of programming languages $|L_y|$ used during an AoC edition, and we show these in Figure~\ref{fig:Language-Participants}.

As expected, many participants usually start with a programming language and continue the competition with the same one. There are very few cases in which participants switch programming languages and use different ones for solving the same puzzle or solving new ones. We conclude that there are very few users that change their programming language during an AoC edition. 

In order to study how participants change programming languages in different AoCs, we first extract the Jaccard coefficients of the possible combinations of sets of participants $V_y$ in the 2019-2021 editions. The results are reported in Table \ref{tab:Participant-Intersection}.

\begin{table}[!ht]
\centering
\begin{tabular}{c|c|c|c}
\toprule
Sets & Intersection & Union & Jaccard \\ \midrule
$V_{2019}$, $V_{2020}$ & 343 & 3,238 & 0.11 \\ 
$V_{2020}$, $V_{2021}$ & 675 & 4,686 & 0.14 \\ 
$V_{2019}$, $V_{2021}$ & 251 & 3,721 & 0.07 \\ 
$V_{2019}$, $V_{2020}$, $V_{2021}$ & 188 & 5,376 & 0.03
\\ \bottomrule
\end{tabular}
\caption{Common participants in the 2019-2021 AoC editions}
\label{tab:Participant-Intersection}
\end{table}

From the analysis of Table \ref{tab:Participant-Intersection}, we observe that some users attend two different editions of AoC. The highest Jaccard coefficient is 0.14 and is computed between the participants of the 2020 and 2021 editions, which means that 14\% of total users attended both. In the three years of AoC we studied, the common participants are few, only 3\% of the total. However, we can still study common participants and deepen our analysis on the behavior behind the change of programming languages during an AoC edition or between two editions. To this end, we first consider the participants who tried to solve at least two puzzles in the same AoC or in two different editions. Formally speaking, we can define the set of users who solved more than one puzzle changing programming languages during a series of time instants as $\tilde{V}_{{t_1}..{t_n}} = \{ v : v \in V, \exists t \in t_2..t_n, |\delta_{t_1}(\{v\}) \bigtriangleup \delta_t(\{v\})| > 0\}$, where $\bigtriangleup$ indicates the symmetric differences between sets. If $v$ changed programming language in a time instant $t \in t_1..t_n$ then the cardinality of the set obtained from the symmetric difference between $\delta_{t_1}(\{v\})$ and $\delta_t(\{v\})$ is higher than 0. If $t_1..t_n = t$ is just a time instant, we can define the set $\tilde{V}_t = \{v : v \in V_t,~|L_{vt}| > 1\}$. This is the set of participants who used more than one language during the same time instant $t$. By definition, given $v$, $L_{vt}$ cannot contain more than one language $l$ if the participant has not written more than 1 comment during $t$. 

From this, we can also define the function $\tau_{intra}(\tilde{V_t}) = \frac{|\tilde{V_t}|}{|V_t|}$ that gives the percentage of participants that changed the programming language during $t$, and the function $\tau_{inter}(\tilde{V}_{t_1..t_n}) = \frac{|\tilde{V}_{t_1..t_n}|}{|\bigcup V_{t}~\forall t \in t_1..t_n|}$ for different time instants $t_1..t_n$. We compute the results for the considered editions and we report them in Table \ref{tab:Change-Same-Challenge} and Table \ref{tab:Change-Different-Challenge}.

\begin{table}[!ht]
\centering
\begin{tabular}{c|c|c|c}
\toprule 
$y$ & $\tau_{intra}(\tilde{V}_y)$ & $1 - \tau_{intra}(\tilde{V}_y)$ & $|\tilde{V}_y|$  \\ \midrule 
2019 & 0.22 & 0.78 & 500 \\ 
2020 & 0.22 & 0.78 & 1,307 \\ 
2021 & 0.20 & 0.80 & 1,561
\\ \toprule
\end{tabular}
\caption{Number of participants changing languages during the same AoC edition}
\label{tab:Change-Same-Challenge}
\end{table}

\begin{table}[!ht]
\centering
\small
\begin{tabular}{c|c|c}
\toprule
$\tau_{inter}(\tilde{V}_{2019..2021})$ & $1-\tau_{inter}(\tilde{V}_{2019..2021})$ & $|\tilde{V}_{2019..2021}|$  \\ \midrule
0.47 & 0.53 & 918
\\ \bottomrule
\end{tabular}
\caption{Number of participants changing languages in different AoC editions}
\label{tab:Change-Different-Challenge}
\end{table}

From the analysis of Table \ref{tab:Change-Same-Challenge}, we note that only 20-22\% of participants change programming language during the same AoC. Different results can be observed in Table \ref{tab:Change-Different-Challenge}, in which 47\% of users change languages between AoCs, while 53\% use the same in all AoC editions.

This difference means that participants who attend more than one AoC edition do not tend to use always the same programming language, while they tend to employ the same programming language during a single AoC edition. This phenomenon could highlight that the AoC initiative is effective since people who participated in previous editions test new programming languages when they participate in the next ones. To test if these differences between the values obtained before during the same AoC and in different AoCs are statistically significant, we run a $\chi^2$ test whose results are reported in Table \ref{tab:Adoption-Chisquare}.

\begin{table}[!ht]
\centering
\begin{tabular}{c|c|c}
\toprule
$y$ & $p$-value & $\chi^2$ \\ \midrule
2019 & 1.67 $e^{-21}$ & 90.70 \\ 
2020 & 7.16 $e^{-35}$ & 151.76 \\ 
2021 & 4.74 $e^{-46}$ & 202.95 \\ \midrule
All years & 9.40 $e^{-55}$ & 253.89
\\ \bottomrule
\end{tabular}
\caption{$\chi^2$ tests between $\tau_{intra}(\tilde{V_t})$ and $1-\tau_{intra}(\tilde{V_t})$, and $\tau_{inter}(\tilde{V}_{2019..2021})$ and $1-\tau_{inter}(\tilde{V}_{2019..2021})$nel}
\label{tab:Adoption-Chisquare}
\end{table}

From Table \ref{tab:Adoption-Chisquare}, we observe that all the $p$-values are below 0.05, which means that the differences in the adoption of programming languages in the same AoC edition and in two AoCs editions are statistically significant. Therefore, we conclude that participants joining the next AoC tend to use new programming languages different from the previous one, so they have the chance to get their hands on new tools. In this perspective, AoC is able to gain the interest of the online community in solving puzzles through programming and motivates frequent participants to test new programming languages.

\subsection{RQ3: Programming language resiliency}
\label{sub:RQ3}

In this section, we deepen our analysis on programming languages. Specifically, we investigate two aspects: {\em (i)} the maximum duration of a programming language, and {\em (ii)} the attractivity of a programming language when a participant decides to change it. First of all, we must study each programming language to characterize them. For this reason, we split the programming languages into two groups according to their programming paradigm: Imperative and Declarative. The languages that allow both paradigms are added to both groups. In order to make this first distinction and add further characteristics, we use the Stack Overflow survey 2022\footnote{\url{https://survey.stackoverflow.co/2022/}}. Stack Overflow is recognized as one of the best communities for programmers and provides many important tips for any task. Stack Overflow also provides an annual survey in which it reports the most used, popular, and loved programming languages, along with information about salaries and employment, and much more. Starting from this survey, we create two further categorizations for each programming language. The first is ``Loved vs Not loved'', where we split the programming languages according to developers' opinions reported in the Stack Overflow survey. Similarly, the second categorization concerns ``Popular vs Not popular''. So, for each programming language, we have three categorizations, which are fundamental for the investigation of the two aspects reported above.

\begin{figure}[!ht]
\centering
\includegraphics[width=0.45\textwidth]{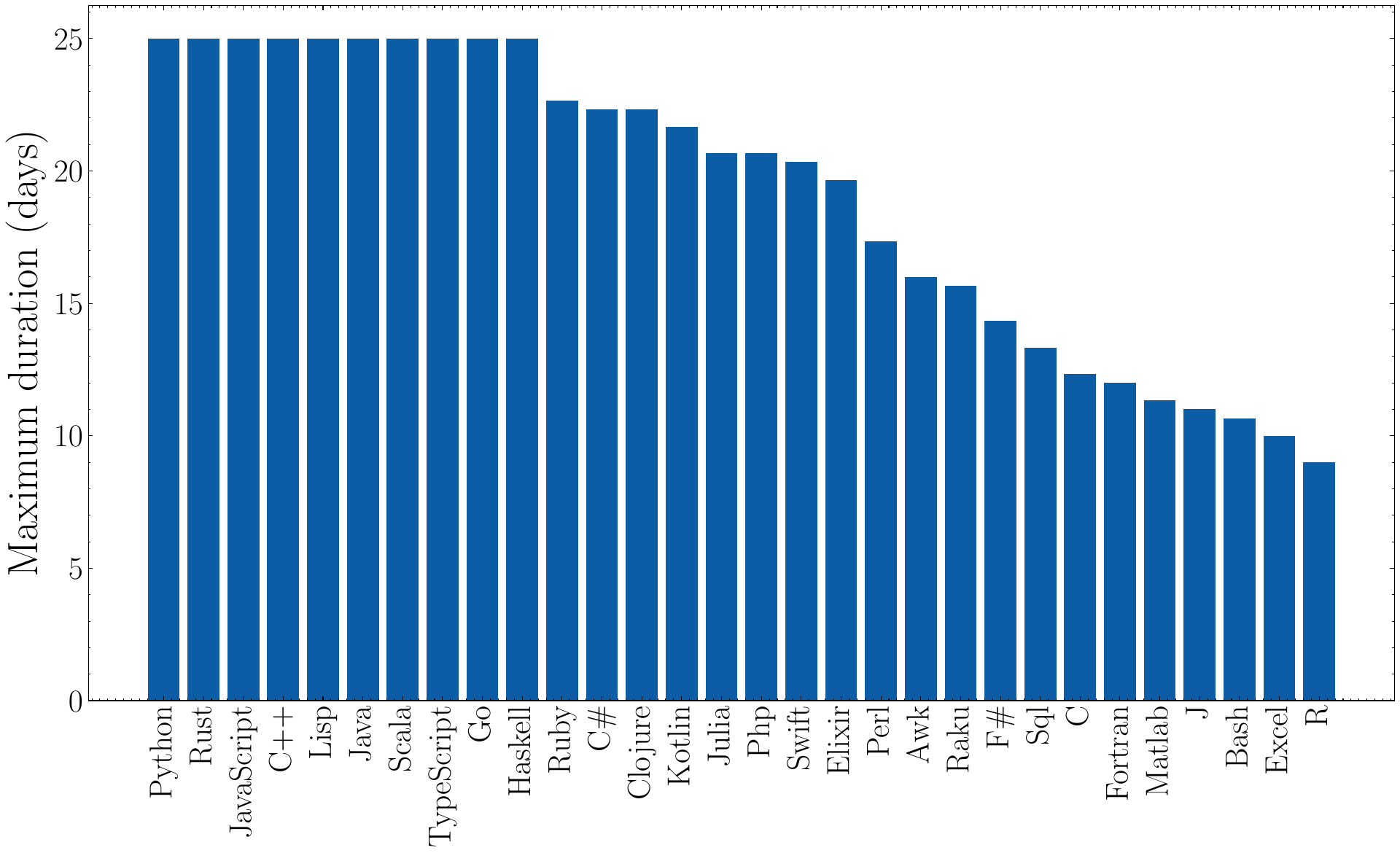}
\caption{Mean of the maximum duration of programming languages in all AoC editions}
\label{fig:Max-Duration}
\end{figure}

We want to first investigate the maximum duration of a programming language $l$, which is defined as $a_{ly} = \max\limits_{v \in V_y} \{ max(A_{vl})\}$. Roughly speaking, $a_{ly}$ represents the maximum number of consecutive days in which the programming language $l$ has been used in the edition $y$. The range of $a_{ly}$ is between 0 ($l$ is not used in any comment) and 25 ($l$ appears in at least a comment on all days). We also define $\bar{a}_l$ as the mean of the maximum duration of a programming language $l$ in all editions and report the corresponding top 30 values in Figure~\ref{fig:Max-Duration}.

From the analysis of Figure \ref{fig:Max-Duration}, we observe that only 10 programming languages out of 30 achieve the highest $\bar{a}_l$ in the 2019-2021 AoC editions. Comparing this figure with Figure \ref{fig:Most-Used-Programming}, these 10 programming languages are also the most cited in comments. Then, $\bar{a}_l$ starts to rapidly decrease so that the last $l$ of the top 30 achieves 9 days.

Now, we use the characterization defined above to verify if there exists a statistically significant difference in the durations of programming languages. To this end, we split the languages according to ``Imperative vs Declarative'', ``Loved vs Not loved'', ``Popular vs Not popular'', compute $\bar{a}_l$ for each $l$, and run Kolmogorov-Smirnov tests between the pairs of distributions. The corresponding results are reported in Table \ref{tab:KS-test-MaxDuration}.

\begin{table}[!ht]
\centering
\begin{tabular}{c|c|c}
\toprule
Comparison & $p$-value & $K$-$S$ statistic \\ \midrule
Imperative vs Declarative & 0.999  & 0.154 \\
Loved vs Not loved & 8.489e-9 & 0.786 \\
Popular vs Not popular & 3.714e-7 & 0.714
\\ \bottomrule
\end{tabular}
\caption{Kolmogorov-Smirnov tests results on the mean of the maximum duration of the programming languages}
\label{tab:KS-test-MaxDuration}
\end{table}

From the analysis of Table \ref{tab:KS-test-MaxDuration}, we surprisingly see that the programming paradigm has no influence on the maximum duration of a programming language. However, there are statistical differences between the categories ``Loved vs Not loved'' and ``Popular vs Not popular''. Indeed, in this case, $\bar{a}_l$ between these pairs of distributions are statistically different, which means that the choice of a programming language that is popular or is loved by the programming community will influence the duration of the attendance of the corresponding participants. Moreover, since the values of $\bar{a}_l$ for all $l$ of ``Loved'' and ``Popular'' (resp., 17.27 and 17.70) are higher than the ones of ``Not loved'' and ``Not popular'' (resp., 2.49 and 2.67), we not only state that these distributions are statistically different, but also that ``Loved'' and ``Popular'' programming languages achieve much higher durations than ``Not loved'' and ``Not popular''. In conclusion, participants are more likely to conclude an AoC edition if the language used is loved or popular, despite its programming paradigm.

\begin{figure}[!ht]
\centering
\includegraphics[width=0.45\textwidth]{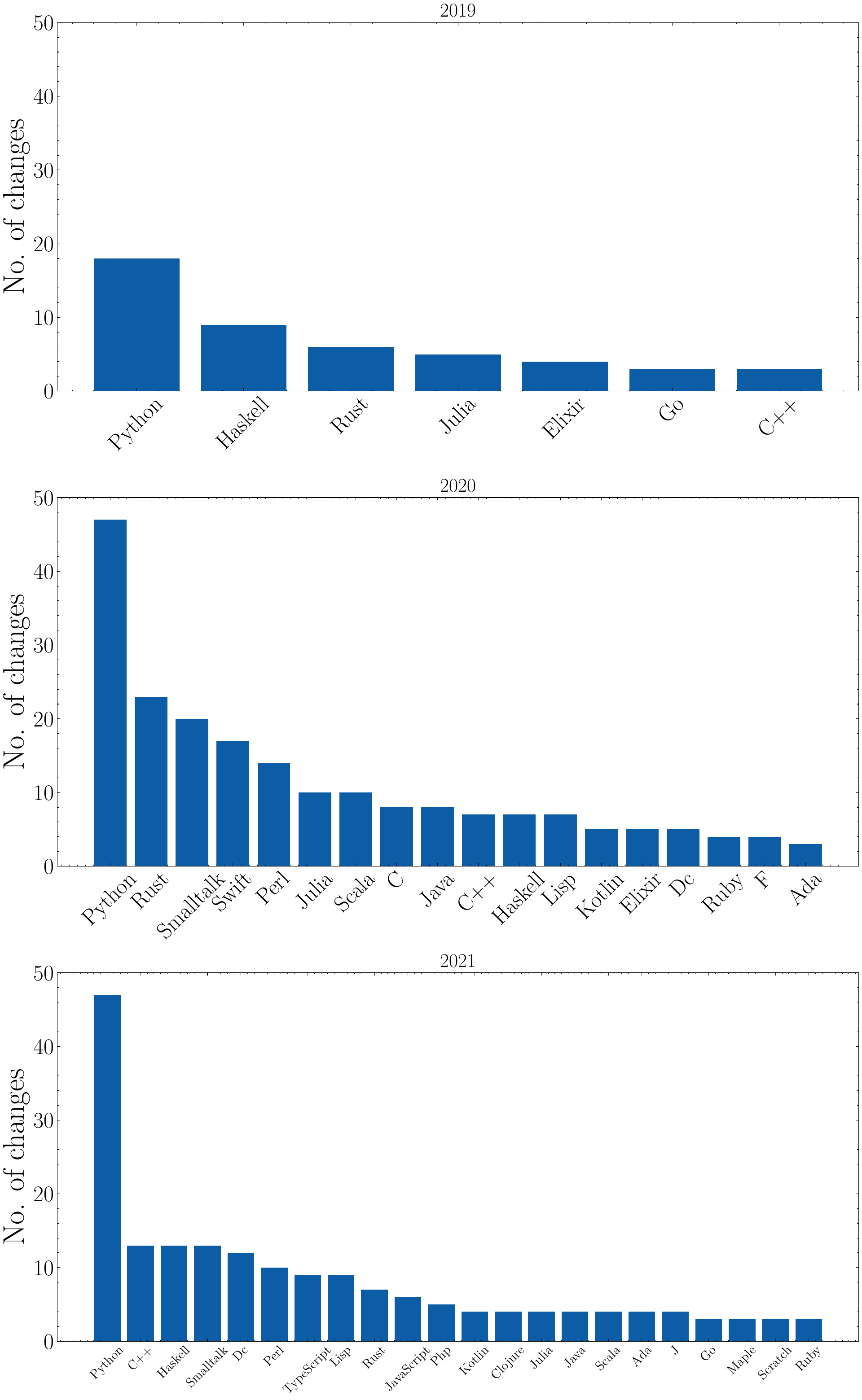}
\caption{Number of times the programming language is chosen for a switch for each AoC edition}
\label{fig:Change-Different-Challenge}
\end{figure}

The second aspect we investigate is the change of programming language during an AoC edition. As we observed in RQ2, there is a percentage of participants that switch languages during the same AoC. Starting from this, it is interesting to study the attractiveness of a programming language, and so how much it influences participants to change language. This switch could be due to complex puzzles (i.e., participants switch back to well-known programming languages), to test new programming languages, or to try to solve a puzzle with different approaches. In order to study this, we consider the scenario in which a participant uses a programming language $l$ at time $t$ and then switches to $l'$ at time $t+1$. We define $c_{l'} = |E_{l'}|$ as the number of times $l'$ is chosen as new language. We show in Figure \ref{fig:Change-Different-Challenge} the distribution of this number for each $l' \in L$ and for each AoC edition.

From the analysis of Figure \ref{fig:Change-Different-Challenge}, we observe that in 2019 there have been very few programming language changes, while this phenomenon is much more relevant in 2020 and 2021. Specifically, Python is the most chosen one, probably for its popularity and its ease of use. Participants may find themselves struggling with certain puzzles and so they choose an easier programming language (such as Python, Rust, C++, or Swift). 

As previously done, we group programming languages in ``Imperative vs Declarative'', ``Loved vs Not loved'', ``Popular vs Not popular'', in order to verify statistically significant differences in the distribution of $c_{l'}$ and, therefore, check if the participants show preferences when switching languages. For each $l' \in L$, we compute the mean of $c_{l'}$ for all the AoC editions and use a Kolmogorov-Smirnov test between the pairs of distributions. The corresponding results are reported in Table \ref{tab:KS-test-Change}.

\begin{table}[!ht]
\centering
\begin{tabular}{c|c|c}
\toprule
Comparison & $p$-value & $K$-$S$ statistic \\ \midrule
Imperative vs Declarative & 0.999 & 0.117 \\
Loved vs Not loved & 4.917e-4 & 0.505 \\
Popular vs Not popular & 5.947e-4 & 0.515
\\ \bottomrule
\end{tabular}
\caption{Kolmogorov-Smirnov tests results on the pairs of distributions of $C_{l}$}
\label{tab:KS-test-Change}
\end{table}

From the analysis of Table \ref{tab:KS-test-Change}, we note that also in this case the programming paradigm does not influence the choice of a new programming language to use. The participant does not seem concerned by the programming style but looks more interested in the programmer community's thoughts. Indeed, we observe statistical differences between the distributions of ``Loved vs Not loved'' and ``Popular vs Not popular''. When a participant decides to switch to a new programming language, he/she will very likely choose one that is ``Loved'' or ``Popular'' according to the programmer community of Stack Overflow. Furthermore, we computed the means of the distributions of `Loved'' and ``Popular'' (resp., 15.12 and 15.24) are higher than ``Not loved'' and ``Not popular'' (resp., 4.37 and 4.38), we point out that ``Loved'' and ``Popular'' programming languages are chosen many more times than the ones of the ``Not loved'' and ``Not popular'' categories.

\section{Related Work}
\label{sec:Related-Work}
There are a few research directions of interest that intersect the context of our study, namely, (i) the analysis of social media contexts for characterizing users' behavior adopting programming languages and (ii) the analysis of programming language adoption and usage.

As for the first direction, there are various studies that can be considered related to ours. In particular, a considerable amount of studies have analyzed the Stack Overflow (SO) platform in order to shed light on the programming languages' adoptions and developers' interactions. For the uninformed reader, SO is a social media platform for questioning and answering (Q\&A) related to programming in general. The seminal work in~\cite{nasehi2012makes} presented a qualitative analysis of the questions and answers posted on SO with the aim of identifying effective examples. This analysis reports that the explanations given with the examples are as important as the examples themselves. In~\cite{wang2013empirical}, the authors focus on developer interactions within the platform, resulting in several insights about how developers approach this kind of platform. The utility of community Q\&A platforms like SO is the object of study of ~\cite{ndukwe2022perceptions}, in which qualitative interview-guided analysis is reported. The analysis highlights the positive and negative experiences of practitioners using such platforms. Other works on the topic focus on the quality of questions~\cite{baltadzhieva2015predicting} and on the reliability of code snippets provided in answers~\cite{zhang2018code}. The Q\&A process on SO is thoroughly investigated in~\cite{zhu2022empirical}.

As for the second direction, the effort produced in the literature focuses on studying the adoption and usage of various programming languages in different contexts. More in detail, several empirical analyses have been presented regarding this aspect~\cite{meyerovich2013empirical,ray2014large,bissyande2013popularity}. In~\cite{meyerovich2013empirical}, large datasets and surveys of programmers are analyzed and the authors report several findings, such as that the adoption of a programming language is mostly related to its ecosystem rather than its intrinsic features. It is worth pointing out how some of these aspects are also important in our context of study. The work in~\cite{ray2014large} focuses on programming languages and code quality in GitHub. Here, a mixed-method study is conducted in order to analyze the effects of language features on software quality. A similar study is presented in~\cite{bissyande2013popularity}, in which aspects such as popularity, interoperability, and impact of programming languages are examined in 100,000 open-source projects. In this work, the authors find that earlier popular languages, such as C, are still in use with a large code base, while other languages, for instance, Javascript, have experienced increasing popularity due to their usage in novel contexts, like web development.

It is worth noting that our study could also have implications in the use of programming competitions and similar educational gamification efforts~\cite{zhan2022effectiveness,rodrigues2021gamification,shahid2019review}. However, such a discussion is out of the scope of this paper.

All the discussed studies are orthogonal to our proposed one. Although AoC presents a large online community of participants, the event itself does not provide a platform for enabling participants' interactions, thus they resort on an external platform, e.g., Reddit. Therefore, being able to identify the interactions between participants in the challenge needs an additional extraction step. This step is generally not needed in the presented studies. Also, the AoC challenge is not directly designed to assess educational metrics. Nevertheless, the content produced by participants' interactions enables several research questions that, in turn, could also be explored in one of the aforementioned contexts. Also, it can shed light on the motivations and experiences of participants, and this information could be used to inform the design of programming challenges and competitions. To the best of our knowledge, this is the first study to focus on the general AoC challenge and to analyze the Reddit content related to it. The analysis of these aspects can provide insights into how people approach and solve programming challenges.

\section{Conclusion}
\label{sec:Conclusion}

In this paper, we investigated the popular coding challenge called Advent of Code. Specifically, we created a dataset from the {\tt /r/adventofcode} subreddit, focusing on specific discussions (i.e., megathreads), where participants comment with their solutions to daily puzzles. We modeled the scenario as a stream graph that allowed us to represent all the actors through time: participants, comments, and programming languages. Then, we investigated some aspects of the AoC challenges, such as user participation, adoption of new languages during a challenge and between two challenges, and resiliency of the programming languages during an AoC edition. For all these aspects, we obtained interesting findings. We figured out that the most commonly used programming languages remain the same during the three AoC challenges, which means that these tools are very important for the programming community. Furthermore, participants tend to keep the same programming language throughout the challenge, while those who participate in two AoCs tend to switch to a new one. Finally, we observed that the programming languages that are deemed ``Popular'' or ``Loved'' tend to be used for longer periods during an AoC. Moreover, when participants decide to switch languages during an AoC, ``Popular'' or ``Loved'' languages are the most commonly chosen.

In the future, we would like to extend our study in several directions. For instance, we plan to study the comment tree that is created after a solution is posted in the megathread. Indeed, the replies of the community could unveil interesting information hidden in the reactions of other users. Furthermore, starting from a specific programming language, we want to investigate the similarities between the posted solutions. In this way, we could analyze if users are leveraging the same patterns or the same reasoning to solve an AoC puzzle.

\section{Broader Impacts and Ethical Considerations}
There are some implications of our work that are ought to be discussed. The insights depicted in this paper could impact educational gamification efforts as well as the design of programming competitions. Nevertheless, our insights could be useful as a foundation for observing how people approach and solve programming challenges. This aspect requires a more in depth study, which is out of the scope of this paper.

As for ethical considerations, we use Reddit as a data source, and in particular the subreddit \texttt{/r/adventofcode}. All data is public and no treatments were conducted by the research team. Reddit data is anonymous in the sense that Reddit users act anonymously and do not share any personal information. Additionally, we only present aggregate information. Figure~\ref{fig:Comment-Example} shows an example of comment; however, its content is fictitious, and therefore it does not pinpoint a specific Reddit user. We did not seek nor attempt to deanonymize any data, thus we do not reveal any private information about users and the content they produce. The dataset we use can be exactly reproduced by using the Reddit API. We encourage the interested reader to contact us for more detail on the dataset analysis.

Finally, our study targets a periodical programming challenge. The participants in this challenge share their intentions on the aforementioned subreddit. Therefore, interested third-parties could push unsolicited ads by posting in the subreddit during the course of the challenge. However, we do not suggest nor condone this act in any means in this study.


\end{document}